\documentstyle[twocolumn,aps,floats,amssymb,epsfig]{revtex}   
\begin{document}
\newcommand{\EGLOB}{E_{\rm glob}}
\newcommand{\ELOC}{E_{\rm loc}}
\newcommand{\ERANDGLOB}{E_{\rm glob}^{\rm random}}
\newcommand{\ERANDLOC}{E_{\rm loc}^{\rm random}}
\newcommand{\be}{\begin{equation}}
\newcommand{\ee}{\end{equation}}
\newcommand{\bea}{\begin{eqnarray}}
\newcommand{\n}{\nonumber\\}
\newcommand{\eea}{\end{eqnarray}}
%
\twocolumn[\hsize\textwidth\columnwidth\hsize\csname @twocolumnfalse\endcsname

\title{\bf Economic Small-World Behavior in Weighted Networks}
\author{ Vito Latora$^{1}$ and Massimo Marchiori$^{2,3}$ }
\address{ $^1$ Dipartimento di Fisica e Astronomia,
               Universit\'a di Catania, and INFN,\\
               Corso Italia 57, I-95129, Catania, Italy\\
          $^2$ W3C and Lab. for Computer Science,
               Massachusetts Institute of Technology,\\
               545 Technology Square, Cambridge, MA 02139, USA\\
         $^3$  Dipartimento di Informatica, Universit\'a di Venezia, Italy }
\date{\today}
\maketitle

\begin{abstract}
The small-world phenomenon has been already the subject of a huge
variety of papers, showing its appeareance in a variety of
systems. However, some big holes still remain to be filled, as the
commonly adopted mathematical formulation 
suffers from a variety of limitations, that make it unsuitable to provide
a general tool of analysis for real networks,  
and not just for mathematical (topological) abstractions. 
In this paper we show where the major problems arise, 
and how there is therefore the need for a new reformulation of 
the small-world concept. 
Together with an analysis of the variables involved, 
we then propose a new theory of small-world networks based on two 
leading concepts: efficiency and cost. 
Efficiency measures how well information
propagates over the network, and cost measures how expensive it is
to build a network. The combination of these factors leads us to
introduce the concept of {\em economic small worlds\/}, that
formalizes the idea of networks that are "cheap" to build, and
nevertheless efficient in propagating information, both at global
and local scale. This new concept is shown to overcome all the
limitations  proper of the so-far commonly adopted formulation,
and to provide an adequate tool to quantitatively analyze the
behaviour of complex networks in the real world. Various complex
systems are analyzed, ranging from the realm of neural networks, to
social sciences, to communication and transportation networks. In
each case, economic small worlds are found. Moreover, using the
economic small-world framework, the construction principles of 
these networks can be quantitatively analyzed and compared, 
giving good insights on how efficiency and economy principles 
combine up to shape all these systems.
\end{abstract}

\vspace{0.5cm}
]     

\section{INTRODUCTION}
\label{sec:intro} 
There is a revolution in the making when it
comes to understanding the complex world around us \cite{yaneer}.
For decades we have been taught to look for the source of all
complex behaviors in the properties of the system's simple
constituents: the main idea to approach a physical problem was
based on the fact that any physical system, even if extreme
complicate, would simplify when studied at smaller and smaller
scale and divided into many simple systems. In the last years this 
view has rapidly changed, with the beginning
of a broad movement of interests and researches on
multidisciplinary problems, and the birth of a new science, the
science of complexity \cite{yaneer,gellmann}. Today, the most
accepted definition of a complex system is that of a system made
by a large number of interacting elements or components whose {\it
collective behavior cannot be simply understood in terms of the
behavior of the components}. To make few examples of complex
systems think of a brain, of a social systems, or of a biological
organism. The simple elements of such systems, the neurons in a
brain, the people in the social system and the cells in the
biological organism are strongly interconnected. Even if we know
many things about a neuron or a specific cell, this does not mean
we know how a brain or a biological system works: any approach
that would cut the system into parts would fail. We need instead
mathematical models that capture the key properties of the entire
ensemble. Only such approaches can success in describing the non
trivial mechanism of how the complex behavior of the whole is
related to the behavior of the parts (this aspect is called
emergence \cite{yaneer}). The general characteristics of a complex
system can be listed as follows:
\\
-- The strong interconnection and interdependence of the parts.
The elements of a complex system interact in a nonlinear way,
and are themselves nonlinear dynamical systems.
\\
-- The existence of a rich structure over several scales.
In space, this property is the definition of a fractal.
In time it means that some process of self-organization
is going on, and then the intrinsic order of a complex system
is dynamic rather than static.

Therefore, chaos and statistical physics are two mathematical
disciplines that have found an intensive application in the study
of complex systems. In fact, a complex system can be modelled as a
{\it network}, where the vertices are the elements of the system
and edges represent the interactions between them: neural
networks, social interacting species, coupled biological and
chemical systems, computer networks or Internet are only few of
such examples.

From one side, scientists have concentrated the attention
in the study of the dynamics of coupled chaotic systems. Since
many things are known about the chaotic dynamics of
low-dimensional non linear systems, a great progress has been
achieved in the understanding of the dynamical behavior of chaotic
systems coupled together in a simple, geometrical regular array
(coupled chaotic maps \cite{kaneko}), or in a completely
random way \cite{bak,sneppen}.

A parallel approach (our paper belongs to this),
focuses instead on the architecture of a
complex system: it concerns with
the study of the connectivity properties of the network.
In fact, the network structure can be as important
as the nonlinear interactions between elements.
An accurate description of the coupling architecture and
a characterization of the structural properties
of the network can be of fundamental importance
also to understand the dynamics of the system.
The questions to answer in this case are: how the
networks look like, and how do they emerge and evolve. 
The research about networks has given rather unexpected
results: in fact the statistical physics is able to
capture the topology of many diverse systems within a
common framework, but this common framework is very different
from the regular array, or the random connectivity,
previously used to model the network of a complex system.
In a recent paper Watts and Strogatz have shown that
the connection {\it topology} of some biological,
technological and social networks is neither completely regular
nor completely random \cite{watts,wattsbook}, but stays somehow in
between these two extreme cases.
These particular class of networks,
named {\it small worlds} in analogy with the concept of
small-world phenomenon developed 30 years ago in social psychology
\cite{milgram}, are in fact highly clustered like regular lattices,
yet having small characteristic path lengths like random graphs.
A pictorial description of this situation is that the
networks' complexity lies at the edge of order and chaos.
The original paper of Watts and Strogatz has triggered
a large interest on the study of the properties of
small worlds (see ref. \cite{newman1} for a recent review).
Researchers have focused their attention on different aspects:
study of the inset mechanism \cite{barrat,watts1,lm1,amaral1},
dynamics \cite{lago} and spreading of diseases on small worlds
\cite{newman2}, applications to social networks 
\cite{newman3,newman4,amaral2} 
and to Internet \cite{barabasi1,barabasi2}.

This paper is about the same definition of the small-world
behavior. We show that the study of a generic complex network
poses new challenges, that can in fact be overcome by using a more
general formalism than the one presented by Watts and Strogatz.
The small-world behavior can be defined in a general and more
physical way by considering how efficiently the information is
exchanged over the network. The formalism we propose is valid both
for unweighted and weighted graphs and extends the application of
the small-world analysis to any complex network, also to those
systems where the euclidian distance between vertices is important
and therefore too poorly described only by the topology of
connections. The results of our study, in part already been
presented in ref.\cite{lm2}, are here extended by the introduction
of a new variable quantifying the cost of the network. The paper
is organized as follows. In Section \ref{sec:loro} we examine the
original formulation proposed by Watts and Strogatz for
topological (unweighted) networks. In Section \ref{sec:noi} we
present our formalism based on the the global and local efficiency
and on the cost of a network: the formalism is valid also 
for weighted networks. Then we introduce and discuss four simple 
procedures (models) to construct unweighted and weighted 
networks. These simple models help to illustrate the concepts of 
global efficiency, local efficiency and cost, 
and to discuss the intricate relationships between these 
three variables. We define an economic small-world 
network  as a low-cost system that communicate efficiently both 
on a global and on a local scale. 
In Section \ref{sec:real} we present a series of
applications to the study of real databases of networks of
different nature, origin and size: 1) neural networks (two
examples of networks of cortico-cortical connections, and an
example of a nervous system at the level of connections between
neurons), 2) social networks (the collaboration network of movie
actors), 3) communication networks (the World Wide Web and the
Internet), 4) transportation systems (the Boston urban
transportation systems).

\section{THE WS FORMULATION}
\label{sec:loro} 
We start by reexamining the ``WS formulation'' of
the small-world phenomenon in topological (relational) networks
proposed by Watts and Strogatz  in ref. \cite{watts}. Watts and
Strogatz consider a generic graph $\bf G$ with $N$ vertices
(nodes) and $K$ edges (arcs, links or connections). $\bf G$ is
assumed to be:

{\it 1) Unweighted.} The edges are not
assigned any a priori weight and therefore are all equal.
An unweighted graph is sometimes called a topological or
a relational graph, because the difference between two edges
can only derive from the relations with other edges.

{\it 2) Simple.} This means that either a couple of nodes is
connected by a direct edge or it is not: multiple edges between
the same couple of nodes are not allowed.

{\it 3) Sparse.} This property means that $K \ll N(N-1)/2$,
i.e. only a few of the total possible number of edges $N(N-1)/2$
exist.

{\it 4) Connected.} $K$ must be small enough to satisfy property
3, but on the other side it must be large enough to assure that
there exist at least one path connecting any couple of nodes. For
a random graph this property is satisfied if $K \gg N \ln N$.

All the information necessary to describe such a graph are
therefore contained in a single matrix $\{a_{ij}\}$, the so-called
{\it adjacency} (or connection) {\it matrix}. This is a $N \cdot N$
symmetric matrix, whose entry $a_{ij}$ is $1$ if there is an
edge joining vertex $i$ to vertex $j$, and $0$ otherwise.
Characteristic quantities of graph $\bf G$, which will be used in the
following of the paper, are the degrees of the vertices. 
The degree of a vertex $i$ is defined as the
number $k_i$ of edges incident with vertex $i$, i.e. the number
of neighbours of $i$. 
The average value of $k_i$ is $k=1/N ~\sum_i k_i = 2K/N$.
In order to quantify the structural properties of $\bf G$, Watts and Strogatz
propose to evaluate two quantities: the characteristic path length $L$ and
the clustering coefficient $C$.

\subsection{The characteristic path length $L$}
\label{length} 
One of the most important quantities to
characterize the properties of a graph is the geodesic, or the
shortest path length between two vertices (popularly known in
social networks as the number of degrees of separation
\cite{milgram,kochen,guare}). 
The shortest path length $d_{ij}$ between $i$ and $j$ 
is the minimum
number of edges traversed to get from a vertex $i$ to another
vertex $j$. By definition $d_{ij} \ge 1$, and $d_{ij}=1$ if there
exists a direct edge between $i$ and $j$. In general the geodesic
between two vertices may not be unique: there may be two or more
shortest paths (sharing or not sharing similar vertices) with the
same length (see ref.\cite{newman3,newman4} for a graphical example of a
geodesic in a social system, the collaboration network of
physicists). The whole matrix of the shortest path lengths
$d_{ij}$ between two generic vertices $i$ and $j$ can be extracted
from the adjacency matrix  $\{a_{ij}\}$ (there is a huge
number of different algorithms in the literature from the standard
breadth-first search algorithm, to more sophisticated 
algorithms \cite{algorithm}). 
The characteristic path length $L$ of graph $\bf G$ is 
defined as the average of the shortest path lengths between 
two generic vertices. 
\be 
L({\bf G})= \frac {1}{N(N-1)} \sum_{i\neq j \in {\bf G} } d_{ij} 
\label{l}
\ee 
Of course the assumption that $\bf G$ is connected (see
assumption number 4) is crucial in the calculation of $L$. It
implies that there exists at least one path connecting any couple
of vertices with a finite number of steps, $d_{ij}$ finite
$\forall i \neq j$, and therefore it assures that also $L$ is a
finite number. For a generic graph (removing the assumption of
connectedness) $L$ as given in eq.(\ref{l}) is an ill defined
quantity, because can be divergent.

\subsection{The clustering coefficient $C$}
\label{clustering} 
An important concept, which comes from social
network analysis, is that of transitivity \cite{wasserman}. In
sociology, network transitivity refers to the enhanced probability
that the existence of a link between nodes (persons or actors)
$i$ and $j$ and between nodes $j$ and $k$, implies the existence
of a link also between nodes $i$ and $k$. In other words in a social
system there is a strong probability that a friend of your friend
is also your friend. The most common way to quantify the
transitivity of a network $\bf G$ is by means of the fraction of
transitive triples, i.e. the fraction of connected triples of
nodes which also form triangles of interactions; this quantity can
be written as \cite{newman3,newman4,newman5}: 
\be 
T({\bf G}) = {
\mbox{ $3 \times$ \# of triangles in {\bf G} }
        \over
       \mbox{ \# of connected triples of vertices in {\bf G} }
     }
\ee 
The factor 3 in the numerator compensates for the fact that
each complete triangle of three nodes contributes three connected
triples, one centered on each of the three nodes, and ensures that
$T=1$ for a completely connected graph \cite{newman3,newman4}. 
As already said,  
$T$ is a classic measure used in social sciences to indicate how much,
locally, a network is clustered (how much it is "small world", so
to say). 
%
\begin{figure}
\begin{center}
\epsfig{figure=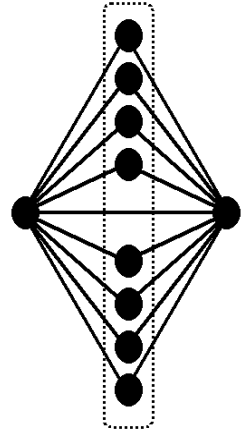,width=1.\columnwidth,angle=0}
\end{center}
\caption{The above network is composed by $N+2$ nodes in total:
$N$ nodes (the ones contained in the dotted square), plus other two
nodes on the two sides. The transitivity $T$ for such network is equal 
to $3/(N+2)$, and therefore becomes zero in the limit of large networks. 
On the other hand, the clustering coefficient $C$ is 
$\frac{N^2+N+4}{N^2+3N+2}$,
which always stays close to one.\label{fig:CisnotT}}
\end{figure}
%
In ref.\cite{watts} Watts and Strogatz use instead
another quantity to measure the local degree of clustering. 
They propose to calculate the so-called clustering coefficient 
$C$. This quantity gives the average cliquishness of the nodes of 
$\bf G$, and is defined as follows. 
First of all a quantity $C_i$, the local clustering coefficient of node
$i$, is defined as: 
\bea 
C_i &=& { \mbox{ \# of edges in $\bf G_i$
}
        \over
        \mbox{maximum possible \# of edges in  $\bf G_i$ }
     }=
\n
    &=& {  \mbox{  \# of edges in  $\bf G_i$ }
        \over
        \mbox{  $k_i(k_i-1)/2$ }
     }
\eea where $\bf {G_i}$ is the subgraph of neighbours of $i$, and
$k_i$ is the number of neighbours of vertex $i$. Then at most
$k_i(k_i-1)/2$ edges can exist in $\bf {G_i}$, this occurring when
the subgraph $\bf {G_i}$ is completely connected (every neighbour
of $i$ is connected to every other neighbour of $i$). $C_i$
denotes the fraction of these allowable edges that actually exist,
and the clustering coefficient $C({\bf G})$ of graph $\bf G$ is
defined as the average of $C_i$ over all the vertices $i$ of $\bf
G$: 
\be C({\bf G}) = \frac {1}{N} \sum_{i \in {\bf G} } C_i 
\ee 
In definitive $C$ is the average cliquishness of the nodes of $\bf
G$.
It is important to observe that $C$, although apparently similar
to $T$, is in fact a different measure. For example, consider the 
network in fig.\ref{fig:CisnotT}: for that network, as $N$ gets
large the transitivity gets worst and worst, and $T$ approaches $0$.
On the other hand, $C$ instead always stays close to $1$.
Therefore, while in many occasions $C$ is indeed a good
approximation of transitivity, it is in fact a totally different
measure. We will see in the rest of the paper how in fact $C$ can
be seen as the approximation of a different measure (efficiency).

\subsection{The small-world behavior: the WS model}
\label{tswb1}
The mathematical characterization of the small-world
behavior proposed by Watts and Strogatz is based on the
evaluation of the two quantities we have just defined:
the characteristic path length $L$, measuring the
typical separation between two generic nodes in the network,
and the clustering coefficient $C$, measuring the average
cliquishness of a node.
As we will see in the following of this
section, small-world networks are somehow in between regular
and random networks: they are highly clustered like regular lattices,
yet having small characteristics path lengths like random graphs.
%
\begin{figure}
\begin{center}
\epsfig{figure=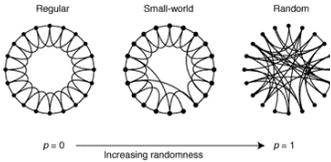,width=0.5\columnwidth,angle=270}
\end{center}
\caption{The rewiring procedure of the WS model interpolates
between a regular lattice and a random graph without altering the
number of nodes or edges. The regular lattice has $N=20$ nodes,
each connected to its $4$ neighbours ($k=4$), and a total number of edges
$K=40$. As the rewiring probability $p$ increases, the network
becomes increasingly disordered. For $p=1$ a random graph is
obtained. After Watts and Strogatz \protect\cite{watts}.
\label{fig:WSmodel}}
\end{figure}
%
In ref.\cite{watts} Watts and Strogatz propose a one-parameter
model (the WS model) to construct a class of graphs $\bf G$ which
interpolates between a regular lattice and a random graph. The WS
model is a method to produce a class of graphs with increasing
randomness without altering the number of nodes or edges: an
example is reported in Fig.\ref{fig:WSmodel}. The WS model starts
with a one-dimensional lattice with $N$ vertices, $K$ edges, and
periodic boundary conditions. Every vertex in the lattice is
connected to its $k$ neighbours ($k=4$ in figure). The random
rewiring procedure consists in going through each of the edges in
turn and independently with some probability $p$ rewire it.
Rewiring means shifting one end of the edge to a new vertex chosen
randomly with a uniform probability, with the only exception as to
avoid multiple edges (more than one edge connecting the same
couple of nodes), self-connections (a node connected by an edge to
itself), and disconnected graphs. In this way it is possible to
tune $\bf G$ in a continuous manner from a regular lattice ($p=0$)
into a random graph ($p=1$), without altering the average number
of neighbours equal to $k = 2K/N$.
%
\begin{figure}
\begin{center}
\epsfig{figure=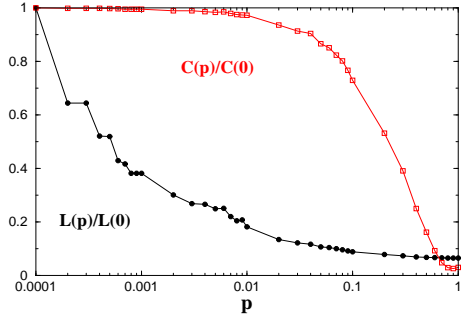,width=0.7\columnwidth,angle=270}
\end{center}
\caption{Characteristic path length $L$, and clustering
coefficient $C$ for the class of graphs produced by the WS model.
Here we consider $N=1000$ and $k=10$. As function of $p$ the
rewiring procedure interpolates between a regular lattice ($p=0$)
and a random graph ($p=1$), and produces the small-world behavior
for $p$ in the range 0.01-0.1\label{fig:CL-graph}}
\end{figure}
%
Let us examine first the behavior of $L$ and $C$ in the two limiting
cases (an analytical estimate is possible in both
cases \cite{watts,barrat,bollobas}):
\\
--- for the regular lattice ($p=0$ in the WS model),
we expect $L\sim N/2k$ and a relatively
high clustering coefficient $C= 3/4 (k-2)/(k-1)$.
\\
--- for the random graph ($p=1$ in the WS model),
we expect $L \sim \ln N /ln(k-1)$ and $C\sim k/N$.
\\
It is worth to stress how regular and random graphs behave
differently when we change the size of the system $N$. If we
increase $N$, keeping fixed the average number of edges per vertex
$k$, we see immediately that for a regular graph $L$ increases
with the size of the system, while for a random graph $L$
increases much slower, only logarithmically with $N$. On the other
hand, the clustering coefficient $C$ does not depend on $N$ for a
regular lattice, while it goes to zero in large random graphs.
From these two limiting cases one could argue that short $L$ is
always associated with small $C$, and long $L$ with large $C$.
Instead social systems, which are a paradigmatic example of a
small-world network, can exhibit, at the same time, short
characteristic path length\cite{milgram} like random graphs, and
high clustering \cite{wasserman}. Now we can come back to the WS
model. To understand the coexistence of small characteristic path
length and high clustering, typical of the small-world behavior,
we report in Fig.\ref{fig:CL-graph}  the behavior of $L$ and $C$
as a function of the rewiring probability $p$ for a graph with
$N=1000$ and $k=10$. We follow the same lines of \cite{watts} and
we normalize the two quantities by their value at $p=0$ in order
to have $0 \le C(p)/C(0) \le 1$ and $0 \le L(p)/L(0) \le 1
~\forall p$. Although in the two limiting cases large $C$ is
associated to large $L$ and viceversa small $C$ to small $L$, the
numerical experiment reveals very interesting properties in the
intermediate regime: only few rewired edges (small $p \neq 0$) are
sufficient to produce a rapid drop of $L$, while $C$ is not
affected and remains equal to the value for the regular lattice.
In this intermediate regime the network is highly clustered 
like regular lattices and has 
small characteristic path lengths like random graphs. These
networks are named small worlds in analogy with the small-world
phenomenon empirically observed in social systems more than 30
years ago by the social psychologist Stanley Milgram
\cite{milgram}. Milgram performed the first experiment to measure
the length of the shortest acquaintance chain between two generic
individuals in United States, and found an average length equal to
5, a value extremely small if compared to the population of the
United States (about 200 millions in 1967). The WS model is a way
to construct networks with the characteristics of a small-world.
Of course the main question to ask now is if the small-world
behavior is only a feature of an abstract model as the WS model,
or if it can be present in real networks. The mathematical
formalism presented can be used to analyze real systems. Watts and
Strogatz have applied their mathematical formalism, based on the
evaluation of $L$ and $C$ to study the topological properties of
real networks databases. In their paper \cite{watts} they consider
three different networks:
\\
1) an example of social network, the collaboration graph
of actors in feature films \cite{actors},
\\
2) the neural network of a nematode, the {\em C.~elegans\/} \cite{verme}
as an example of a biological network
\\
3) a technological network, the electric power grid of the western
United States.
\\
They show that the three networks, when considered as
unweighted networks, are all examples of small worlds.

\section{A NEW FORMULATION VALID FOR WEIGHTED NETWORKS}
\label{sec:noi} Having a mathematical characterization of the
small-world behavior makes it tempting to apply the same concept
to any complex system. This grand plan clashes with the fact that
the mathematical formalism of Watts and Strogatz suffers from
severe limitations.
\\
First of all it works only in the {\it topological abstraction}
(the approximation of unweighted network), where the only
information retained is about the existence or the absence of a
link, and nothing is known about the physical length of the link
(or more generically the weight associated to the link, see the
first assumption in the original formulation of Section
\ref{sec:loro} where the graph {\bf G} is assumed to be
unweighted), and multiple edges between the same couple of nodes
are not allowed (see the second assumption: {\bf G} must be
simple).  
\\
Moreover it applies only to some cases, whereas in general the two
quantities $L$ and $C$ are ill-defined: for example the assumption
number four of connectedness (see Section \ref{sec:loro}) is
necessary because otherwise the quantity $L$ would diverge.

The inadequacy of the Watts and Strogatz formalism is already
evident to a more accurate analysis of the same three examples
presented in their paper. Let us analyze the three examples one by
one. In the case of films actor two are the problems: the original
formalism can not be applied directly to the whole network, but it
works only when the analysis is restrained to the giant connected
component of the graph\cite{watts} in order to avoid the
divergence of $L$. Moreover the topological approximation only
provides whether actors participated in some movie together, or if
they did not at all. Of course, in reality there are instead
various degrees of correlation: two actors that have done ten
movies together are in a much stricter relation than two actors
that have acted together only once. We can better shape this
different degree of friendship by using a non-simple graph or by
using a weighted network: if two actors have acted together we
associate a weight to their connection by saying that the length
of the connection, instead of being always equal to one, is equal
to the inverse of the number of movies they did together. In the
case of the neural network of the {\em C.~elegans\/} Watts and
Strogatz define an edge in the graph when two vertices are
connected by either a synapse or a gap junction\cite{watts}. This
is only a crude approximation of the real network. Neurons are
different one from the other, and some of them are in much
stricter relation than others: the number of junctions connecting
a couple of neurons can vary a lot, up to a maximum of 72. As in
the case of film actors a weighted network is more suited to
describe such a system and can be defined by setting the length of
the connection $i-j$ as equal to the inverse number of junctions
between $i$ and $j$. To conclude with the last example presented
by Watts and Strogatz, the electrical power grid of the western
United States, which is clearly a network where the geographical
distances play a fundamental role. Any of the high voltage
transmission lines connecting two stations of the network has a
length, and the topological approximation of the Watts and
Strogatz's mathematical formalism, which neglect such lengths, is
a poor description of the system. Of course a generalization of
the original formalism to weighted networks would allow the study
of the connectivity properties of many complex systems, extending
the application of the small-world concept to a realm of new
networks previously not considered. A very significative example
is that of a transportation system : public transportation (bus,
subway and trains), highways, airplane connections. Transportation
systems can be analyzed at different levels and in this paper we
will present an example of an application to urban public
transportation.

The problems in the passage from abstract networks
to real complex systems can be overcome by using a more general
formalism, in part already presented in ref. \cite{lm2},
and here described in details and extended by the
introduction of a new variable quantifying the cost of
a network. In the following of this Section we show that:

1) A weighted network can be characterized by introducing the
variable {\em efficiency\/} $E$, which measures how efficiently
the nodes exchange information.
The definition of small-world behavior can be formulated in terms
of the efficiency:
this single measure evaluated on a global and on a local scale
plays in turn the role of $L$ and $C$.
Small-world networks result as systems that are both
globally and locally efficient.

2) The formalism is valid both for weighted and unweighted
(topological) networks. In the case of topological networks our
formalism does not coincide exactly with the one given
by Watts and Strogatz. For example our formalism applies to
unconnected graphs.

3) An important quantity, previously not considered is the
{\it cost} of a network. Often high (global and local) efficiency
implies an high cost of the network.

We are now ready to describe our new formalism.
Since in general a real complex system is better described
by a weighted network, we now start by considering a
generic graph $\bf G$ as a {\it weighted} and
possibly even {\it non-connected} and
{\it non-sparse} graph.
A weighted graph needs two matrices to be described:

\noindent
-- the {\it adjacency matrix} $\{a_{ij}\}$, containing the
information about the existence or not existence of a link, and
defined as for the topological graph as a set of numbers
$a_{ij}=1$ when there is an
edge joining $i$ to $j$, and $a_{ij}=0$ otherwise;

\noindent -- a matrix of the weights associated to each link. We
name this matrix $\{\ell_{ij}\}$ the {\it matrix of physical
distances} because the number $\ell_{ij}$ can be imagined as the
space distance between $i$ and $j$. We suppose $\ell_{ij}$ to be
known even if in the graph there is no edge between $i$ and $j$.
To make a few concrete examples:
\\
$\ell_{ij}$ can be identified with the geographical distance
between stations $i$ and $j$ both in the case of the electrical
power grid of the western United States studied by Watts and
Strogatz, and in the case of other transportation systems
considered in this paper. In such a situation $\ell_{ij}$ respect
the triangular inequality though in general this is not a
necessary assumption.
\\
The presence of multiple edges, typical of the neural network of
the {\em C.~elegans\/} and of social systems like the network of
films actors, can be included in the same framework by setting
$\ell_{ij}$ equal to the inverse number of edges between $i$ and
$j$ (respectively the inverse number of junctions between two
neurons, or the inverse of the number of movies two actors did
together). This allows to remove the hypothesis of simple network
in the (assumptions number 2 in the formalism of of Watts and
Strogatz) and to consider also $\it non-simple$ systems as
weighted networks. The resulting weighted network is, of course, a
case in which the triangular inequality is not satisfied. For a
computer network or Internet $\ell_{ij}$ can be assumed to be
proportional to the time needed to exchange a unitary packet of
information between $i$ to $j$ through a direct link. Or as
$1/v_{ij}$, the inverse velocity of a chemical reactions along a
direct connection in a metabolic network. Of course, in the
particular case of an unweighted (topological) graph $\ell_{ij}=1
~\forall i \neq j$.

\subsection{The efficiency $E$}
\label{} 
In a weighted graph, the definition of the shortest path
length $d_{ij}$ between two generic points $i$ and $j$, is
slightly different than the definition used in Section
\ref{sec:loro} for an unweighted graph. In this case the shortest
path length $d_{ij}$ is in fact defined as the smallest sum of the
physical distances throughout all the possible paths in the graph
from $i$ to $j$. Again, when $\ell_{ij}=1 ~\forall i \neq j$, i.e.
in the particular case of an unweighted graph, $d_{ij}$ reduces to
the minimum number of edges traversed to get from $i$ to $j$.
\\
The matrix of the shortest path lengths $\{d_{ij}\}$ is therefore
calculated by using the information contained both in matrix 
$\{a_{ij}\}$ and in matrix $\{\ell_{ij}\}$ \cite{algorithmnostro}.
We have $d_{ij} \ge \ell_{ij} ~\forall i,j$, the equality being
valid when there is an edge between $i$ and $j$. Let us now
suppose that every vertex sends information along the network,
through its edges. We assume that the efficiency $\epsilon_{ij}$
in the communication between vertex $i$ and $j$ is inversely
proportional to the shortest distance:  $\epsilon_{ij} = 1/d_{ij}
~\forall i,j$. Note that here we assume that efficiency and
distance are inversely proportional. This is a reasonable approximation 
in general, and in particular for all the systems
considered in this paper. 
Of course, sometimes other relationships might be used, especially 
when justified by a more specific knowledge about the system. 
By assuming $\epsilon_{ij} = 1/d_{ij}$, 
when there is no path in the graph 
between $i$ and $j$ we get $d_{ij}=+\infty$ and consistently
$\epsilon_{ij}=0$. Consequently the average {\it efficiency} of
the graph $\bf G$ can be defined as \cite{performance1}:
\begin{equation}
\label{efficiency}
E({\bf G})=
\frac{ {{\sum_{{i \ne j\in {\bf G}}}} \epsilon_{ij}}  } {N(N-1)}
          = \frac{1}{N(N-1)}
{\sum_{{i \ne j\in {\bf G}}} \frac{1}{d_{ij}}}
\end{equation}
Throughout this paper we consider {\it undirected} graphs, i.e.
there is no associated direction to the links. This means that
both  $\{\ell_{ij}\}$ and $\{d_{ij}\}$ are symmetric matrices and
therefore the quantity $E({\bf G})$ can be defined simply by using
only half of the matrix as: $E({\bf G})= \frac{2}{N(N-1)}
{\sum_{{i<j\in {\bf G}}} \frac{1}{d_{ij}}}$. Anyway we prefer to
give the more general definition (\ref{efficiency}) since our
formalism can be easily applied to directed graphs as well.

Formula (\ref{efficiency}) gives a value of $E$ that can vary in
the range $[0,\infty[$. It would be more practical to have $E$
normalized to be in the interval $[0,1]$. $E$ can be normalized by
considering the ideal case ${\bf {G^{ideal}}}$ in which the graph
$\bf G$ has all the $N(N-1)/2$ possible edges. In such a case the
information is propagated in the most efficient way since $d_{ij}
=\ell_{ij}~\forall i,j$, and $E$ assumes its maximum value $E({\bf
{G^{ideal}}})= \frac{1}{N(N-1)} {\sum_{{i \ne j\in {\bf G}}}
\frac{1}{\ell_{ij}}}$. The efficiency $E({\bf G})$ considered in
the following of the paper are always divided by $E({\bf
{G^{ideal}}})$ and therefore $0 \le E({\bf G}) \le 1$. Though the
maximum value $E=1$ is typically reached only when there is an
edge between each couple of vertices, real networks can
nevertheless assume high values of $E$.

\subsection{Global and local efficiency}
\label{} One of the advantages of the efficiency-based formalism
is that a single measure, the efficiency $E$ (instead of the two
different measures $L$ and $C$ used in the WS formalism) is
sufficient to define the small-world behavior.

In fact, on one side, the quantity defined in equation
(\ref{efficiency}) can be evaluated as it is for the whole
graph $\bf G$ to characterize the {\it global efficiency}
of $\bf G$.
We therefore name it $\EGLOB$:
\begin{equation}
\label{}
 \EGLOB =  \frac {E({\bf G})}  {E({\bf G^{ideal}})}
\end{equation}
\noindent
As said before, the normalization factor ${E({\bf G^{ideal}})}$
is the efficiency of the ideal case ${\bf {G^{ideal}}}$
in which the graph $\bf G$ has all the $N(N-1)/2$ possible edges.
Being the efficiency in communication between two generic vertices,
$\EGLOB$ plays a role similar to the inverse of the characteristic
path length $L$. In fact $L$ is the mean of $d_{ij}$, while $\EGLOB$ is
the average of $1/d_{ij}$, i.e. the inverse of the harmonic mean 
of $\{d_{ij}\}$. 
Nowadays the harmonic mean finds extensive
applications in a variety of different fields: 
in particular it is used to calculate
the average performance of computer systems\cite{performance1,benchmarkart},
parallel processors\cite{performance4}, and
communication devices (for example modems and
Ethernets\cite{performance2}). In all such cases, where a mean flow-rate
of information has to be computed, the simple arithmetic mean
gives the wrong result. 
As we will see in Section (\ref{compare}) and in Section (\ref{tswb2}),
in some cases $1/L$ gives a good approximation of $\EGLOB$,
although $\EGLOB$ is the real variable to be considered when we want
to characterize the efficiency of a system transporting
information in parallel.
In the particular case of a disconnected graph the difference between
the two quantities is evident because $L=+\infty$ while
$\EGLOB$ is a finite number.

On the other side the same measure, the efficiency, can be
evaluated for any subgraph of $\bf G$, and therefore it can be
used also to characterize the local properties of the graph. In
the WS formalism it is not possible to use the characteristic path
length for quantifying both the global and the local properties of
the graph simply because $L$ can not be calculated locally, most
of the subgraphs of the neighbors of a generic vertex $i$ being
disconnected. In our case, since $E$ is defined also for a
disconnected graph, we can characterize the local properties of
$\bf G$  by evaluating for each vertex $i$ the efficiency of $\bf
{G_i}$, the subgraph of the neighbors of $i$. We define the {\it
local efficiency} as:
\begin{equation}
\label{}
\ELOC = 1/N  \sum_{i \in {\bf G}} ~  \frac{E(\bf {G_i})} {E(\bf {G_i^{ideal}})}
\end{equation}
\noindent
Here, for each vertex $i$, the normalization factor
${E({\bf G_i^{ideal}})}$ is the efficiency of the ideal case
${\bf {G_i^{ideal}}}$ in which the graph $\bf G_i$ has all the
$k_i(k_i-1)/2$ possible edges.
$\ELOC$ is an average of the local efficiency and
plays a role similar to the clustering coefficient $C$.
Since $i \notin \bf {G_i}$, the local efficiency $\ELOC$ tells how much
the system is {\it fault tolerant}, thus how efficient
is the communication between the first neighbours of $i$ when $i$
is removed.
This concept of fault tolerance is different from
the one adopted in Ref. \cite{barabasi3,cohen,lm5},
where the authors consider the response of the entire
network to the removal of a node $i$.
Here the response of the subgraph of first neighbours of $i$
to the removal of $i$ is considered.

We can now introduce a new, generalizing, definition of
small-world, built in terms of the characteristics of information
flow at global and local level: a {\it small-world network\/} is a
network with high $\EGLOB$ and $\ELOC$, i.e. very efficient both
in global and local communication. This definition is valid both
for unweighted and for weighted graphs, and can also be applied to
disconnected graphs and/or non sparse graphs.

\subsection{Comparison between $\EGLOB, \ELOC$ and $L, C$}
\label{compare} 
\begin{figure}
\begin{center}
\epsfig{figure=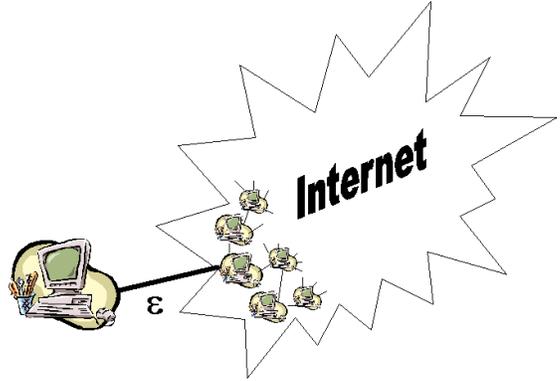,width=.9\columnwidth,angle=0}
\end{center}
\caption{We attach a new computer to the Internet (which
already had $N$ nodes), with a connection represented by a 
small efficiency $\varepsilon$. 
Having one (or few) computer with an extremely slow connection, 
does not mean that the whole Internet diminishes by far its efficiency: 
in practise, the presence of such slow computer goes unnoticed, 
because the other thousands of computers are exchanging 
packets among them in a very efficient way. 
$L$ fails to properly capture the global behaviour of
systems like the Internet, 
unlike $\EGLOB$, that perfectly matches the observed behaviour (see text). 
\label{fig:internet}}
\end{figure}

It is interesting to study more in detail the
correspondence between our measure and the quantities $L$ and $C$
of \cite{watts} (or, correspondingly, $1/L$ and $C$). The
fundamental difference is that $1/L$ measures the efficiency of a
{\em sequential system\/}, that is to say, of a system where there
is only one packet of information going along the network. On the
other hand, $\EGLOB$ measures the efficiency for {\em parallel
systems\/}, where all the nodes in the network concurrently
exchange packets of information. This can explain why $L$ works
reasonably: it can be seen that $1/L$ is a reasonable
approximation of $\EGLOB$ when there are not huge differences
among the distances in the graph, and so considering just one
packet in the system is more or less equivalent to the case where
multiple packets are present. This is the case for all the
networks presented in \cite{watts}, and this effect is
strengthened even more by the fact that the topology only is
considered. Having explained why $L$ behaves relatively well in
some case, it is also worth noticing that, like every
``approximation'', it fails to properly deal with all cases. For
example, the sequentiality of the measure $1/L$ explains why many
limitations have to be introduced, like connectedness, that are
present just in order to make the formulas valid. Consider the
limit case where a node is isolated from the system. In the case
of a neural network, this corresponds for example to the death of
a neuron. In this case, $1/L$ drops to zero ($L=+\infty$), which
is of course not the overall efficiency of the system: in fact,
the brain continues to work, as all the other neurons continue to
exchange information; only, the efficiency is just slightly
diminished, as now there's one neuron less. And, correctly, this
is properly taken into account using $\EGLOB$. Even without
dropping the connectedness assumption, another example can show
how in the limit case, the approximation given by $1/L$ diverges
from the real efficiency measure. 
Let us consider the Internet and the 
situation represented in Figure \ref{fig:internet}: 
suppose we attach a new computer 
to the Internet (which already had $N$ nodes), 
with efficiency $\varepsilon$, that can be
seen as the speed of the connection. 
This happens every time the Internet is augmented with a new computer, 
and every time we turn on our computer in the office. 
A situation like this occurs daily in the order of the millions. 
How does it globally affect the Internet, 
according to $L$ and $\EGLOB$? 
It can be proved that $L$ augments by approximately
$\frac{1}{\varepsilon(N+1)}$. This means that if for any reason,
the connection speed is particularly slow (or becomes such, for
example due to a congestion, or the computer gets low in
resources), the whole Internet's $L$ is heavily affected and can
rapidly become enormous. Even, whenever the computer blocks (or
it's shut down), $L$ diverges to infinity (like, so to say, if the
Internet had collapsed).
On the other hand, the efficiency $\EGLOB$ has a relative
decrement of approximately $\frac{2}{N+1}$, which means that in
practice, as $N$ is quite large, the particular behaviour of the new
computer affects the Internet in a negligible way.
Summing up, having one or few computer with an extremely slow connection, 
does not mean that the whole Internet
diminishes by far its efficiency: in practise, the presence of
such few very slow computers goes unnoticed, because the
other thousands of computers are exchanging packets among them in
a very efficient way.  
Therefore, $L$ fails to properly capture the global behaviour of
systems like the Internet ($1/L$ would give a number very
close to zero because, it measures the average efficiency in case
a single packet is active thorough the Internet), 
unlike $\EGLOB$, that perfectly matches the observed behaviour. 
%
%
The crucial point here is the following: 
all the networks considered in
\cite{watts} to justify the definition of small-worlds (and, in
fact, most of the networks the model complex systems) are {\em
parallel systems\/}, where all the nodes interact in parallel
(Internet, World Wide Web, social networks, neural systems and so
on). With this assumption, $\EGLOB$ measures the real efficiency
of the system, and $1/L$ is just a first rough approximation, as
it deals with the sequential case only.

We turn now our attention to $C$ and $\ELOC$. As we have seen in
Section \ref{sec:loro}, the true meaning of the clustering
coefficient $C$ cannot be sought in the classic clustering measure
of social sciences, i.e. transitivity: the two quantities may
diverge, giving diametrically opposite results for the same
networks. On the other hand, it can be shown that $C$, in the case of
undirected topological graphs, is always a reasonable
approximation of $\ELOC$. Therefore, the seemingly ad-hoc nature
of $C$ in the WS formalism, now finds a new meaning in the general
notion of efficiency: there are not two different kinds of
properties to consider when analyzing a network on the local and
on the global scale, but just one unifying concept: the
efficiency to transport information.

\subsection{The Cost of a Network}
\label{}
An important variable to consider, especially when we
deal with weighted networks and when we want to analyze and
compare different real systems, is the cost of a network.
In fact, we expect the efficiency of a graph
to be higher as the number of edges in the graph increases.
As a counterpart, in any real network there is a price to pay
for number and length (weight) of edges.
In particular the 'short cuts', i.e. the rewired edges that
produce the rapid drop of $L$ and the onset of the small-world
behavior in the WS model connect at no cost vertices that
would otherwise be much farther apart.
It is therefore crucial to consider weighted networks and
to define a variable to quantify the cost of a network.
In order to do so, we define the {\it cost} of the
graph $\bf G$ as:
\begin{equation}
Cost({\bf G})= \frac {\sum_{{i \ne j\in {\bf G}}}
a_{ij}\gamma(\ell_{ij})}
                     {\sum_{{i \ne j\in {\bf G}}} {\gamma(\ell_{ij})}}
\end{equation}
Here, $\gamma$ is the so-called {\em cost evaluator} function,
which calculates the cost needed to build up a connection with a 
given length. 
Of course, $\gamma$ could be equivalently defined on efficiencies 
rather than distances (so, indicating in a sense the cost to set
up a communication channel with the given efficiency). Note that
we have already included in the numerator of this definition the
cost of ${\bf G}^{ideal}$, the ideal graph in which all the
possible edges are present. Because of such a normalization, the
$\gamma$ function needs only to be defined up to a multiplicative
constant, and the quantity $Cost({\bf G})$ is defined in the
interval $[0,1]$, assuming the maximum value $1$ for ${\bf
G}^{ideal}$, i.e. when all the edges are present in the graph.
$Cost({\bf G})$ reduces to the normalized number of edges
$2K/N(N-1)$ in the case of an unweighted graph (for example the WS
model).
\\
Unless otherwise specified, we will assume in the following that
$\gamma$ is defined as the identity function: $\gamma(x)=x$. In fact 
such a cost evaluator works for unweighted networks, and also for most
of the real networks, those where the cost of a connection is
proportional to its length (to the euclidean distance for example): 
in all such cases the definition of the cost reduces to 
$Cost({\bf G})=  
({\sum_{{i \ne j\in {\bf G}}}a_{ij} \ell_{ij}}) /
({\sum_{{i \ne j\in {\bf G}}} {\ell_{ij}}})$. 
A different definition of the cost evaluator function  
will be used instead when we represent networks 
with multiple edges as weighted graphs 
(for examples in the weighted {\em C.\ elegans\/} and in the 
weighted movie actors). 

With our formalism based on the two efficiencies $\EGLOB$ and
$\ELOC$, and on the variable $Cost$, all defined in the range from
$0$ to $1$, we can study in an unified way unweighted
(topological) and weighted networks. We therefore define the
following key notion: let us call {\em economic\/} every network
with low $Cost$; then, an {\em economic small-world\/} is a
network having high $\ELOC$ and $\EGLOB$, and low $Cost$ (i.e.,
both economic and small-world).

\subsection{The economic small-world behavior}
\label{tswb2}
 We are now ready to illustrate the three quantities
$\EGLOB$, $\ELOC$ and $Cost$ at work in some practical examples. 
Starting from the original WS model, and proceeding with different
models, we will illustrate how these three quantities behave in a
dynamic environment where the network changes, have some
nontrivial interaction among each other, and give birth to
small-worlds \cite{algorithm,algorithmnostro}.
Model 1 (the WS model) is a procedure to construct a 
family of unweighted networks with a fixed cost. 
Model 2 is a way to construct unweighted networks, this time with 
increasing cost. 
Model 3 and model 4 are two examples of weighted networks. In particular 
in model 4 the length of the edge connecting two nodes is the
euclidean distance between the nodes.

\begin{figure}
\begin{center}
\epsfig{figure=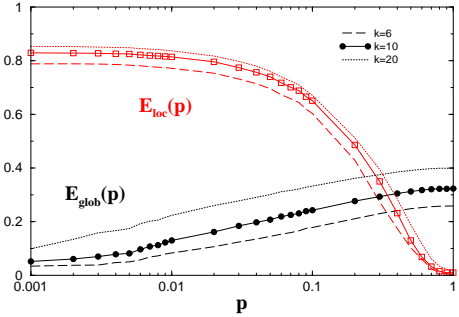,width=0.7\columnwidth,angle=270}
\end{center}
\caption{ Global and local efficiency for model 1 (the WS model),
the class of topological graphs considered by Watts and Strogatz.
A regular lattice with $N=1000$ and $k$ edges per node is rewired
with probability $p$. The logarithmic horizontal scale is used to
resolve the rapid increase in $\EGLOB$ due to the presence of
short cuts and corresponding to the onset of the small-world.
During this increase, $\ELOC$ remains large and almost equal to
the value for the regular lattice. Small worlds have high $\EGLOB$
and $\ELOC$. We consider three different values
$k=6,10,20$ corresponding respectively to $Cost=0.006, 0.01, 0.02$. 
\label{fig:EforWSmodel}}
\end{figure}

\begin{figure}
\begin{center}
\epsfig{figure=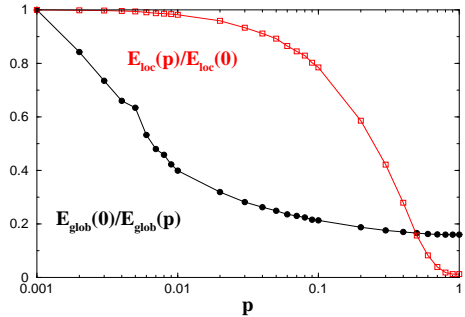,width=0.7\columnwidth,angle=270}
\end{center}
\caption{ Model 1 (the WS model). 
A regular lattice with $N=1000$ and $k=10$ edges per node is rewired
with probability $p$. 
Reporting the quantities $\left(
\frac{\EGLOB(p)}{\EGLOB(0)}\right)^{-1}$ and
$\frac{\ELOC(p)}{\ELOC(0)}$ as a function of $p$,  
the two curves show a behavior similar respectively 
to $L(p)$ and $C(p)$.\label{fig:ElikeLC}}
\end{figure}

{\bf Model 1)} The original WS model is {\em unweighted\/}
(topological): this means we can set $\ell_{ij}=1 ~\forall i \neq
j$, and the quantities $d_{ij}$ reduce to the minimum number of
edges to get from $i$ to $j$. The dynamic changes of the network
consist in rewirings: since the weight is the same for all edges,
also for rewired edges, this means that the $Cost$ (that is
proportional to the total number of edges $K$) does not change
with the rewiring probability $p$. In fig.\ref{fig:EforWSmodel} we
consider a regular lattice with $N=1000$ and three different
values of $k$ ($k=6, 10, 20$), corresponding to networks with
different (low) cost (respectively $Cost=0.006, 0.01, 0.02$), 
and we report $\EGLOB$ and $\ELOC$ as a
function of $p$ \cite{algorithm}. For $p=0$ we expect the system
to be inefficient on a global scale (an analytical estimate gives
$\EGLOB \sim k/N~log(N/K)$), but locally efficient. The situation is
inverted for random graphs. In fact, for example in the case
$k=20$, at $p=1$ $\EGLOB$ assumes a maximum value of $0.4$,
meaning $40\%$ the efficiency of the ideal graph with an edge
between each couple of vertices. This happens at the expenses of
the fault tolerance ($\ELOC \sim 0$). The (economic) small-world
behavior appears for intermediate values of $p$. It results from
the fast increase of $\EGLOB$ caused by the introduction of only a
few rewired edges (short cuts), which on the other side do not
affect $\ELOC$. For the case $k=20$, 
at $p \sim 0.1$, $\EGLOB$ has almost reached the
maximum value of $0.4$, though $\ELOC$ has only diminished by very
little from the maximum value of $0.82$. For such an unweighted
case the description in terms of network efficiency is similar to
the one given by Watts and Strogatz. In fig.\ref{fig:ElikeLC} we
show that if we report the quantities $1/\EGLOB(p)$ and $\ELOC(p)$, 
and we use a normalization similar to the one adopted 
by Watts and Strogatz, i.e. $\EGLOB(0)/\EGLOB(p)$ and
$\ELOC(p)/\ELOC(0)$, we get curves with qualitatively the same
behavior of the curves $L(p)/L(0)$ and $C(p)/C(0)$ (compare with 
fig.\ref{fig:CL-graph}). 
%
%
%

\begin{figure}
\begin{center}
\epsfig{figure=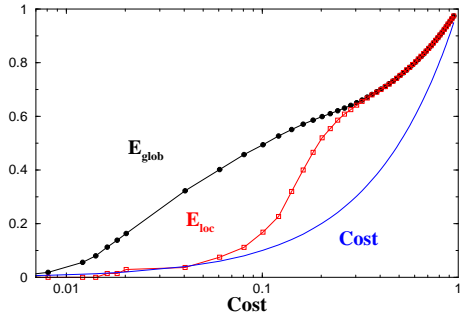,width=0.7\columnwidth,angle=270}
\end{center}
\caption{Model 2. A network is created by adding links 
randomly to an initial configuration with $N=100$ nodes and no links. 
$\EGLOB$ and $\ELOC$ are plotted as functions of the $Cost$. 
The identity curve $Cost$ is also reported to help the reader 
since a logarithmic horizontal scale is used.
\label{easySW}}
\end{figure}

{\bf Model 2)} The above model has proved successful in order to
produce small-worlds, i.e. networks with high $\EGLOB$ and 
high $\ELOC$. However, if that is the goal, then there are
much simpler procedures that can output a small world, even
starting from an arbitrary configuration. For example in 
Fig.\ref{easySW} we consider a model where, starting from a 
configuration with $N=100$ nodes and no links 
we keep adding links randomly, until we reach a  
completely connected network. This model is unweighted 
as model 1. Contrarily to the case of model 1,  
the network changes by adding links, then 
the cost is not a fixed quantity but varies in a monotonic
way, increasing every time we add a link.  
As we can see, for $Cost\sim 0.5-0.6$ we obtain a 
small-world network with $\EGLOB= \ELOC = 0.8$. 
So, if this trivial method manages to produce
small worlds, why can't we find many small worlds like these in
nature ? 
The obvious answer is that here, we are obtaining a
small-world at the expense of the cost: with rich resources (high
cost), the small-world behaviour always appears. In fact, in the limit 
of the completely connected network ($Cost=1$)  
we have $\EGLOB= \ELOC = 1$. 
But what also matters in nature is also economy of a network, 
and in fact a trivial technique like this fails to produce 
{\em economic\/} small worlds.
\\
Note also that the relationship of the variable cost with respect 
to the other two variables is not that trivial. 
Even in the very simple and rigid "monotonic" setting
dictated by this model we observe an interesting behavior of 
the variables $\EGLOB$ and $\ELOC$ as functions of $Cost$. 
In particular we observe a rapid rise of $\ELOC$ when the cost 
increases from $0.1$ to $0.2$. This means that moving from $Cost=0.1$ 
to $Cost=0.2$ we can increase the local efficiency of the network 
from $\ELOC=0.1$ to $\ELOC=0.6$. We therefore obtain a network 
with $60\%$ of the efficiency of the ideal network both on a global 
and local scale, with only the $20\%$ of the cost: this is an 
example of an economic small-world network. 
The effect we have observed has an higher probability to happen 
in the mid-area inbetween the areas of low cost and high cost, 
and it is a first
sign that complex interactions do occur, but not with very low
cost or with very high cost (where economic small-worlds can't be
found).
%
%

\begin{figure}
\begin{center}
\epsfig{figure=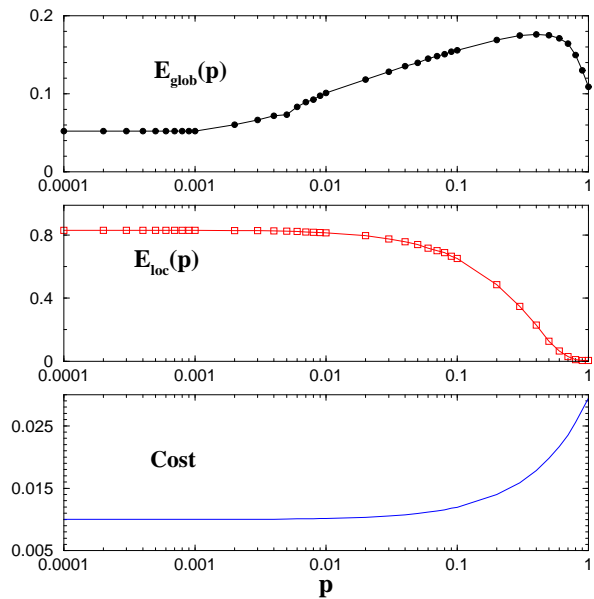,width=0.9\columnwidth,angle=270}
\end{center}
\caption{The three quantities $\EGLOB$, $\ELOC$ and $Cost$
are reported as functions of $p$ in model 3. 
We start with a regular lattice with $N=1000$ and $k=10$
and we implement the same rewiring procedure as in the WS model,
with only difference that the length of the rewired edge is set to
change from the value 1 to the value 3. The economic small-world
behavior shows up for $p\sim 0.1$\label{fig:rewiring-triple}}
\end{figure}

{\bf Model 3)} In this third model, we combine features of the
previous models 1 and 2: we adopt rewiring as in model 1, 
monotonic increase of the cost as in model 2. 
So, while in model 1 the short cuts connect at no 
cost (because $\ell_{ij}=1 ~\forall i \neq j$)
vertices that would otherwise be much farther apart (which is a
rather unrealistic assumption for real networks), in this model
each rewiring has a cost. 
In fig.\ref{fig:rewiring-triple} we implement a random rewiring in
which the length of each rewired edge is set to change from 1 to
3. So, note that this model, unlike the previous two, is {\em
weighted\/}. The figure shows that the small-world behaviour is
still present even when the length of the rewired edges is larger
than the original one. For $p$ around the value $0.1$ we observe
that $\EGLOB$ has almost reached the maximum value $0.18$ ($18\%$
of the global efficiency of the ideal graph with all couples of
nodes directly connected with edges of length equal to $1$) while
$\ELOC$ has not changed too much from the maximum value $0.8$
(assumed at $p=0$). The only difference with respect to model 1 is
that the behaviour of $\EGLOB$ is not simply monotonic increasing.
Of course in this model the variable $Cost$ increases with $p$.
It is interesting to notice that the curve $Cost$ as a function of
$p$, plotted in the bottom of the figure, is specular to the curve
$\ELOC$ as a function of $p$. This means that in the small-world
situation, the network is also economic, in fact the $Cost$ stays
very close to the minimum possible value (assumed of course in the
regular case $p=0$). We have checked the robustness of the results
obtained by increasing even more the length of the rewired edges.
\\
Therefore, this model shows that to some extent, the structure of
a network plays a relevant role in the economy.  Also,
note that in this more complex (weighted) model, behaviour of
$\ELOC$ and $\EGLOB$ become more complex as well: now, $\EGLOB$ is
not a monotonic function of the cost any more, and $\ELOC$ is monotonic,
but decreasing. So, introduction of the weighted model
further shows how the relative behaviour of the three variables
$\EGLOB$, $\ELOC$ and $Cost$ is far from simple.

%
%

\begin{figure}
\begin{center}
\epsfig{figure=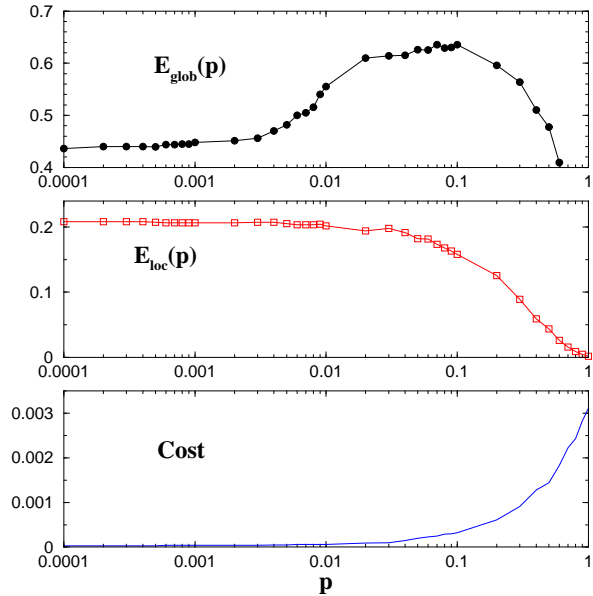,width=0.9\columnwidth,angle=270}
\end{center}
\caption{The three quantities $\EGLOB$, $\ELOC$ and $Cost$
are reported as functions of $p$ in model 4. 
We start with a regular lattice with $N=1000$ and a total
number of edges $K=1507$ (see detail in the text) and we implement
the rewiring procedure with probability $p$. The economic
small-world behavior shows up for $p\sim 0.02-0.04$
\label{fig:rewiring-circle}}
\end{figure}

{\bf Model 4)} As a final example we build on model 3, and ground
it more in reality using a real geometry, in order to investigate
further whether the above effects can also appear in real networks
which are not just mathematical possibilities. In this weighted
model, the length of the edge connecting two nodes is the
euclidean distance between the nodes. The nodes can be placed with
different geometries. Here we consider the case in which the $N$
nodes are placed on a circle as in fig.\ref{fig:WSmodel}. Now the
geometry is important because the physical distance between node
$i$ and $j$ ($i,j=1,...,N$) is defined as the euclidean distance
between $i$ and $j$. In the case of nodes on a circle we have: 
\begin{equation}
\label{metric}
\ell_{ij}= \frac { 2 \sin( |i-j| \pi /N) }
                 { 2 \sin(\pi/N)               }
\end{equation}
In this formula we have set the length of the arc between two
neighbours to be equal to $1$, i.e. $\ell_{ij}= 1$ when
$|i-j|=1$. The radius of the circle is then $R = \frac{1}{2} \sin(
\pi /2) /   \sin(\pi/N)$. In fig.\ref{fig:rewiring-circle} we
report the results obtained by implementing a rewiring procedure
similar to the one considered in the previous models. The only
difference with respect to the previous case is that now we
cannot start from a lattice with $N=1000$, $k=10$. Such a network,
in fact, when considered with the metrics in formula (\ref{metric})
would have $K=5000$ edges and a too high global efficiency, about
$99\%$ of the ideal graph. On the other side, considering as a
starting network a lattice with $k=2$ would affect the local
efficiency. Then we proceed as follows. We create a regular
network with $N=1000$ and $k=6$ and then we eliminate randomly the
$50\%$ of the $3000$ edges to decrease the global efficiency: in
the random realization reported in figure we are left with
$K=1507$ edges. At this point we can implement the usual rewiring
process on this network. For $p\sim 0.02-0.04$ we observe that
$\EGLOB$ has almost reached its maximum value $0.62$ while $\ELOC$
has not changed much from the maximum value $0.2$ (assumed at
$p=0$). As in model 3 the behaviour of $\EGLOB$ is not simply
monotonic decreasing, and as in model 3 the small-world network is
also an economic network, i.e. the $Cost$ stays very close to the
minimum possible value (assumed of course for $p=0$).

So, this model and model 3 suggest that the economic small-world
behavior is not only an effect of the topological abstraction but
can also be found in all the weighted networks where the physical
distance is important and the rewiring has a cost (and, shows how
intricate the relative behaviour of $\EGLOB$, $\ELOC$ and $Cost$
can be).

%
%

\section{APPLICATIONS TO REAL NETWORKS}
\label{sec:real} With our formalism based on the three quantities
$\EGLOB$, $\ELOC$, and  $Cost$, all defined in the range from $0$
to $1$, we can study in an unified way unweighted (topological)
and weighted networks, and we are therefore well equipped to
consider some empirical examples. In this paper we present a study
of 1) neural networks (two examples of networks of
cortico-cortical connections, and an example of a nervous system
at the level of connections between neurons), 2) social networks
(the collaboration network of movie actors), 3) communication
networks (the World Wide Web and the Internet), 4) transportation
systems (the Boston urban transportation systems).

\subsection{Neural Networks}
\label{}
The brain is the most complex and fascinating information
transportation system. Its staggering complexity is the evolutionary
result of adaptivity, functionality and economy.
The brain complexity is already reflected in the complexity
of its structure~\cite{koch}.
Of course neural structures can be studied at several levels of scale.
In fact, thanks to recent experiments, a wealth of neuroanatomical
data ranging from the fine structure of connectivity between single neurons
to pathways linking different areas of the cerebral cortex is now
available.
Here we focus first on the analysis of the neuroanatomical
structure of cerebral cortex, and then on a simple nervous system at
the level of wiring between neurons.
\\
{\bf 1) Networks of Cortico-cortical connections.} The anatomical
connections between cortical areas and group of cortical neurons
are of particular importance because they are considered to have
an intricate relationship with the functional connectivity of the
cerebral cortex \cite{sporns}. We analyze two databases of
cortico-cortical connections in the macaque and in the cat
\cite{scannell1}. The databases consist of the wiring diagrams of
the two system, and there is no information about the weight
associated to the links: therefore we will study these systems as
unweighted networks. The macaque database contains $N=69$ cortical
areas and $K=413$ connections (see ref.\cite{young}, cortical
parcellation after \cite{felleman}, except auditory areas which
follow ref. \cite{pandya}). The cat database has instead $N=55$
cortical areas (including hippocampus, amygdala, entorhinal cortex
and subiculum) and $K=564$ (revised database and cortical
parcellation from \cite{scannell2}). The results in the first two
lines of table \ref{table1} indicates the two networks are
economic small-worlds: they have high global efficiency
(respectively $52\%$ and $69\%$ the efficiency of the ideal graph)
and high local efficiency ($70\%$ and $83\%$ the ideal graph),
i.e. high fault tolerance ~\cite{fault} with only $18\%$ and
$38\%$ of the wirings. Moreover $\EGLOB$ is similar to the value
for random graphs, while $\ELOC$ is larger than $\ELOC^{random}$.
These results indicate that in neural cortex each region is
intermingled with the others and has grown following a perfect
balance between cost, local necessities (fault tolerance) and
wide-scope interactions.
\\
{\bf 2) A network of connections between neurons.}
As a second example we consider the neural network of {\em C.~elegans\/}
the only case of a nervous system completely mapped
at the level of neurons and chemical synapses \cite{white}.
The database we have considered, is the same considered by
Watts and Strogatz and is taken from ref.\protect\cite{verme}.
%
\begin{table}
\caption{The macaque and cat cortico-cortical connections
\protect\cite{scannell1} are two unweighted networks with
respectively $N=69$ and $N=55$ nodes, $K=413$ and $K=564$
connections. Global efficiency, local efficiency and cost are
reported in the first two lines of the table. The results are
compared to the efficiency of random graphs. The nervous system of
{\em C.~elegans\/} is better described by a weighted network: the
network consists of $N=282$ nodes and $K=2462$ edges which can be
of two different kind, either synaptic connections or gap
junctions. This time, associated to each link, there is weight
(see text). In the third line of the table we report the result
for the {\em C.~elegans\/} considered as unweighted (to compare
with cortico-cortical networks), while in the fourth line we
consider the weights. All these systems are examples of economic
small worlds. \label{table1}} 
\begin{tabular}{l|ll|ll|l}
{\em Unweighted:\/} & $\EGLOB$ & $\ERANDGLOB$ & $\ELOC$ & $\ERANDLOC$ & $Cost~~$\\
\tableline
Macaque                    & 0.52 & {\em0.57} & 0.70 & {\em0.35} & 0.18\\
Cat                        & 0.69 & {\em0.69} & 0.83 & {\em0.67} & 0.38\\
\tableline
C. elegans                 & 0.46 & {\em0.48} & 0.47 & {\em0.12} & 0.06\\
\end{tabular}
\begin{tabular}{l|l|l|l}
{\em Weighted:\/} & $\EGLOB$ & $\ELOC$ & $Cost~~$\\
\tableline
C. elegans      & 0.35  & 0.34 & 0.18\\
\end{tabular}
\end{table}
As already discussed in Sect.\ref{sec:noi}, the nervous system of
{\em C.~elegans\/} is better described by a weighted network. In
fact the {\em C.~elegans\/} is a multiple edges system, i.e. there
can be more than one edge (up to $72$ edges) between the same
couple of nodes $i$ and $j$. The presence of multiple edges can be
expressed in our weighted networks formalism by considering a
simple but weighted graph, and setting $\ell_{ij}$ equal to the
inverse number of edges between $i$ and $j$. In this way we get a
weighted network consisting of $N=282$ nodes and $K=2462$ edges
(an edge $i-j$ is defined by the presence of at least one synaptic
connection or gap junction).
Now, observe that doing this choice to weight the system, we then 
have to define appropriately the cost evaluator function $\gamma$
(which can not be the identity any more): the correct choice is to
set $\gamma(x)=1/x$, that is to say, the cost of a connection is
the number of synaptic connections and gap junctions that make it.
\\
In order to compare the {\em C.~elegans\/} to the two
cortico-cortical connections networks, we first consider it as an
unweighted network neglecting the information contained in
$\{\ell_{ij}\}$ (as if $\ell_{ij}=1 ~ \forall i \neq j$).
Similarly to the two cortico-cortical connections networks, the
unweighted {\em C.~elegans\/} is also an economic small-world
network. In third line of table \ref{table1} we see that with a
relative low cost ($6\%$ of the wirings), {\em C.~elegans\/} 
achieves about
a $50\%$ of both the global and local efficiency of the ideal
graph (see also the comparison with the random graph). Moreover
the value of $\EGLOB$ is similar to $\ELOC$. This is a difference
from cortex databases, where fault tolerance is slightly
privileged with respect to global communication. Finally we can
consider the {\em C.~elegans\/} in all its completeness, i.e. as a
weighted graph. Of course in this case the random graph does not
give any more the best approximation for $\EGLOB$. Nevertheless
the values of $\EGLOB$, $\ELOC$ and $Cost$ have a meaning by
themselves, being normalized to the case of the ideal graph. 
We get (see the fourth line of \ref{table1}) that
the {\em C.~elegans\/} is also an economic small-world when
considered as a weighted network with about $35\%$ of the global
and local efficiency of the ideal graph, obtained with a cost 
of $18\%$. It is interesting to notice that, as in
the unweighted case, the system has similar values of $\EGLOB$ and
$\ELOC$ (that is, it behaves globally in the same way as it
behaves locally).
\\
The connectivity structure of the three neural networks
studied reflects a long evolutionary process driven by
the need to maximize global efficiency and to develop
a robust response to defect failure (fault tolerance).
All this at a relatively low cost, i.e. with a small
number of edges, or with a minimum amount of the
length of the wirings.

\subsection{Social Networks}
\label{}
As an example of social networks we study
the collaboration network of movie actors
extracted from the Internet Movie Database\cite{actors},
as of July 1999. The graph considered has $N=277336$ and $K=8721428$,
and is not a connected graph.
The approach of Watts and Strogatz
cannot be applied directly and they
have to restrict their analysis
to the giant connected component of the graph\cite{watts}.
Here we apply our small-world analysis directly to the whole graph,
without any restriction.
Moreover the unweighted case only provides whether actors
participated in some movie together, or if they did not at
all. Of course, in reality there are instead various degrees of
correlation: two actors that have done ten movies together are
in a much stricter relation rather than two actors that
have acted together only once. As in the case of {\em C.~elegans\/} we
can better shape this different degree of friendship by using a
weighted network: we set the distance $\ell_{i,j}$ between two
actors $i$ and $j$ as the inverse of the number of movies they did
together.

As in the case of the {\em C.\ elegans\/}, together with this
choice to weight the system, we also have to define appropriately
the cost evaluator function $\gamma$: the correct choice is
(again) to set $\gamma(x)=1/x$, that is to say, the cost of a
connection between two persons is the number of movies they did
together. 

The numerical values in table \ref{table2} indicate that both the unweighted
and the weighted network shows the economic small-world
phenomenon. In both cases, cost comes out as a leading principle:
this is due somehow to physical limitations, as it is not easy for
actors to perform in a huge number of movies, and for most of
them,  their career is in any case limited in time, while the
database spans all the temporal age. 
%
\begin{table}
\caption{The collaboration network of movie actors (extracted from
the Internet Movie Database, IMD) can be described by an
unweighted or a weighted graph with $N=277336$ and $K=8721428$.} 
\label{table2}
\begin{tabular}{l|ll|ll|l}
{\em Unweighted:\/}   & $\EGLOB$ & $\ERANDGLOB$ & $\ELOC$ &$\ERANDLOC$ & $Cost~~$  \\
\hline
Movie Actors &    0.37&       0.41       &  0.67   &       0.00026     & 0.0002   \\
\end{tabular}

\begin{tabular}{l|l|l|l}
{\em Weighted:\/}    & $\EGLOB$ & $\ELOC$ & $Cost~~$  \\
\hline
Movie Actors   &  0.29   &   0.52   &    0.0005\\
\end{tabular}

\end{table}
%
Of course other social systems can be studied by means
of our formalism: for example the collaboration network
of physicists \cite{newman3,newman4}, the
collaboration network of Marvel comics characters
\cite{marvel}, or some other databases of
social communities \cite{amaral2,newman6}.

\subsection{Communication Networks}
\label{} 
Communication networks
are ubiquitous nowadays: the so-called "information society"
heavily relies on such networks to rapidly exchange information in
a distributed fashion, all over the world. Here, we consider the
two most important large-scale communication networks present
nowadays: the World Wide Web and the Internet. Note that despite
these two networks are often confused and identified, they are
fundamentally different: the World Wide Web (WWW) network is based
on {\em information abstraction\/}, via the fundamental concept of
URI (Uniform Resource Identifier); so, it is not a physical
structure, but an abstract structure. On the other hand, the
Internet is a physical communication network, where each link and
node have a physical representation in space. So, despite these
two communication networks share lot of commonalities (last but
not least, the fact the WWW essentially relies on the Internet
structure to work), they are bottom-down deeply different: one
network (WWW) is purely conceptual, the other one (the Internet)
is physical.
%
\begin{table}
\caption{~~~ Communication networks. Data on the World Wide Web
 from   http://www.nd.edu/\symbol{126}networks
contains $N=325729$ documents and $K=1090108$ links
\protect\cite{barabasi1}, while the Internet database is taken
from http://moat.nlanr.net and has $N=6474$ nodes and $K=12572$
links. Both systems are studied as unweighted graphs and are 
examples of economic small worlds.}
\label{table3}
\begin{tabular}{l|ll|ll|l}
             & $\EGLOB$ & $\ERANDGLOB$ & $\ELOC$ &$\ERANDLOC$ & $Cost~~$  \\
\hline
WWW          & 0.28   &    0.28      &  0.36   &  0.000001  & 0.00002   \\
Internet     & 0.29   &    0.30      &  0.26   &  0.0005    & 0.006     \\
\end{tabular}

\end{table}
%
We have studied a database of the World Wide Web with $N=325729$ documents 
and $K=1090108$ links, and a network of Internet with $N=6474$ nodes 
and $K=12572$ links. Both networks are considered as unweighted 
graphs. 
In table \ref{table3} we report the result of the efficiency-cost 
analysis of the two networks. As we can see, they have relatively high
values of $\EGLOB$ (slightly smaller than the best possible values
obtained for random graphs) and $\ELOC$, together with a very
small cost: therefore, both of them are economic small-worlds.
Observe that interestingly, despite the WWW is a virtual network
and the Internet is a physical network, at a global scale they
transport information essentially in the same way (as their
$\EGLOB$'s are almost equal). At a local scale, the larger $\ELOC$
in the WWW case can be explained both by the tendency in the WWW
to create Web communities (where pages talking about the same
subject tend to link to each other), and by the fact that many
pages within the same site are often quickly connected to each
other by some root or menu page. As far as the cost is concerned,
it is striking to notice how economic these networks are (for
example, compare these data with the corresponding ones for the
cases of neural networks). This clearly indicates that economy is 
a fundamental construction principle of the Internet and of
the WWW.

\subsection{Transportation Networks}
\label{}
\begin{figure}
\begin{center}
\epsfig{figure=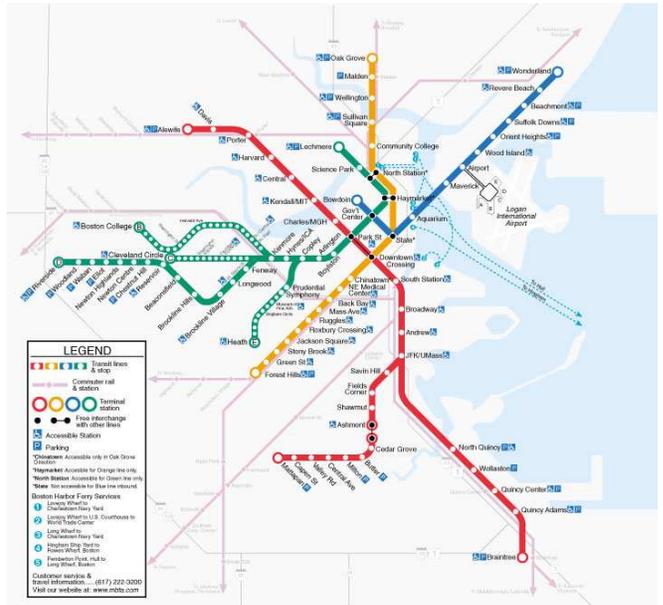,width=1.\columnwidth,angle=0}
\end{center}
\caption{The Boston subway transportation system ({\em MBTA\/})
consists of four different lines (Green, Red, Blue and Orange),
$N=124$ stations and $K=124$ tunnels extending throughout Boston
and the other cities of the Massachusetts Bay
\protect\cite{tboston}.\label{fig:mbta}}
\end{figure}

We focus now on another example of man-made networks, the
transportation networks. As a paradigmatic example of a system
belonging to this class we consider the Boston public
transportation system. Other examples, like the Paris subway systems
and the network of airplanes and highway connections throughout the
world, are currently under study and will be presented in a future
work \cite{lm6}.
\\
The Boston subway transportation system ({\em MBTA\/}) reported in
fig.\ref{fig:mbta} is the oldest subway system in the U.S.\ (the
first electric streetcar line in Boston, which is now part of the
MBTA Green Line, began operation on January 1, 1889) and consists
of $N=124$ stations and $K=124$ tunnels (connecting couples of
stations) extending throughout Boston and the other cities of the
Massachusetts Bay \cite{tboston}. As some of the previous
databases, this is another example of a network better described
by a weighted graph: in this case the matrix $\{\ell_{ij}\}$ is
given by the euclidean distance between $i$ and $j$, i.e. by the
geographical distances between stations. In this sense the {\em
MBTA\/} is a weighted network more similar to the electrical power
grid of the western United States than to weighted networks
representing multiple edges systems like the neural network of the
{\em C.~elegans\/} or to the network of films actors. In fact in
the case of the {\em MBTA\/} the quantities $\ell_{ij}$ respect
the triangle inequality and the definition of the ideal graph is
straightforward since the spatial distance $\ell_{ij}$ between
stations $i$ and $j$ is perfectly defined, independently from the
existence or not of the edge $i-j$. In particular the matrix
$\{\ell_{ij}\}$ has been calculated by using information databases
from the {\em MBTA\/} \cite{tboston},  from the Geographic Data
Technology (GDT), and the U.S.\ National Mapping Division.
%
%
The {\em MBTA\/}, even when considered as an unweighted network,
is a typical example of a case where the WS formalism fails to
apply. We therefore proceed step by step: we first study the
system in the unweighted approximation (illustrating 
that the WS formalism based on $L$ and $C$ does not work, and must
be replaced by the efficiency-based formalism). We finally
represent and study the efficiency of the {\em MBTA\/} in its
completeness, as a weighted network \cite{lm2,lm3}. 
\\
Let us start by showing that even in the approximation of
unweighted network the case of {\em MBTA\/} cannot be considered
by the original formalism, and the efficiency-based formalism must
be used. In the unweighted network approximation the information
contained in $\{\ell_{ij}\}$ is not used (as if $\ell_{ij}=1 ~
\forall i \neq j$). Now, consider for example $L$: if we apply to
the {\em MBTA\/} the original formalism presented in Sect.
\ref{sec:loro}, valid for unweighted (topological) networks, we
obtain $L=15.55$ (an average of 15 steps, or 15 stations to
connect 2 generic stations). And now, to decide if the {\em
MBTA\/} is a small world we have to compare the obtained $L$ to
the respective values for a random graph with the same $N$ and
$K$. But, when we consider a random graph we get $L=\infty$. So,
we are unable to draw any conclusion.
\\
On the other side, the same unweighted network can be perfectly
studied by using the efficiency formalism of Sect. \ref{sec:noi}.
The problem of the divergence we had for $L$ is here avoided,
because when there is no path in the graph between $i$ and $j$,
$d_{i,j}=+\infty$ and consistently $\epsilon_{ij}=0$. The results
are reported in the first line of table \ref{table4} and compared
with the values obtained for the random graph with same number of
$N$ and $K$ (as said before, in the unweighted case, the random
graph provides the best value of $\EGLOB$). We see immediately
that the unweighted network is not a small world because the
$\ELOC$ should be much larger than $\ERANDLOC$, and is instead
smaller than $\ERANDLOC$.
%
\begin{table}
\caption{~~~ The {\em MBTA\/} can be considered as
a network of $N=124$ nodes and $K=124$ links.
The {\em MBTA\/} is first studied as
an unweighted network and then as a weighted network.
Finally the weighted network consisting in the underground
transportation system plus the bus transportation system is
considered as a more complete transportation system.
The matrix $\{\ell_{ij}\}$ has been calculated by using
databases from the {\em MBTA\/} \protect\cite{tboston} and
the U.S. National Mapping Division.
\label{table4}}
\begin{tabular}{l|ll|ll|l}
{\em Unweighted:\/} & $\EGLOB$ & $\ERANDGLOB$ & $\ELOC$ & $\ERANDLOC$ & $Cost~~$\\
\tableline
MBTA    & 0.10  & {\em 0.14} & 0.006  & {\em 0.015} & 0.016\\
\end{tabular}

\begin{tabular}{l|l|l|l}
{\em Weighted:\/} & $\EGLOB$ & $\ELOC$ & $Cost~~$\\
\tableline
MBTA    & 0.63   & 0.03   & 0.002\\
MBTA + bus& 0.72  & 0.46    & 0.004\\
\end{tabular}
\end{table}
In the second line of  table \ref{table4} we report the results
for the weighted case, i.e. the case in which the link
characteristics (lengths in this case) are properly
taken into account, and not flattened into their topological
abstraction.
As a main difference from the unweighted case considered before,
in a weighted case the random graph does not give
the estimate of the highest global efficiency.
In any case the quantities $\EGLOB$ and $\ELOC$
have a meaning by themselves because of the adopted normalization:
the numbers shows {\em MBTA\/} is a very efficient transportation
system on a global scale but not at the local level.
In fact $\EGLOB=0.63$ means that {\em MBTA\/} is only $37\%$
less efficient than the ideal subway with a direct tunnel
from each station to the others. On the other hand
$\ELOC=0.03$ indicates a poor local efficiency:
differently from a neural network or from a social system
the {\em MBTA\/} is not fault tolerant and a damage in a station
will dramatically affect the efficiency in the connection between
the previous and the next station.
To understand better the difference with respect to the other systems
previously considered we need to make few general considerations
about the variable $Cost$ and the rationales in the construction
principles.
As said before in general the efficiency of a graph increases with
the number of edges. As a counterpart, in any real network
there is a price to pay for number and length (weight) of edges. 
If we calculate the cost of the weighted {\em MBTA\/} we get $\it
{Cost}=0.002$, a value much smaller than the ones obtained for
example for the three neural networks considered, respectively
$\it {Cost}=0.18,0.38,0.06-0.07$. This means that {\em MBTA\/}
achieves the $63\%$ of the efficiency of the ideal subway with a
cost of only the $0.2\%$. The price to pay for such low-cost high
global efficiency is the lack of fault tolerance. The difference
with respect to neural networks comes from different needs and
priorities in the construction and evolution mechanism. A neural
network is the results of perfect balance between global and local
efficiency. On the other side, when we build a subway system, the
priority is given to the achievement of global efficiency at a
relatively low cost, and not to fault tolerance. In fact a
temporary problem in a station can be solved in an economic way by
other means: for example, waling, or taking a bus from the
previous to the next station. That is to say, the {\em MBTA\/} is
not a {\em closed system\/}: it can be considered, after all, as a
subgraph of a wider transportation network. This property is very
often so understood that it isn't even noted (consider for
example, the case of the brains), but it is nevertheless of
fundamental importance when we analyze a system: while global
efficiency is without doubt the major characteristic, it is {\em
closure\/} that somehow leads a system to have high local
efficiency (without alternatives, there should be high
fault-tolerance). The {\em MBTA\/}  is not a closed system, and
thus this explains why, unlike in the case of neural networks
fault tolerance is not a critical issue. Changing the {\em MBTA\/}
network to take into account, for example the bus system, indeed,
this extended transportation system comes back to be an economic
small-world network. In fact the numbers in the third line of
table \ref{table4} indicate that the extended transportation system
achieve high global but also high local efficiency ($\EGLOB=0.72$,
$\ELOC=0.43$), at a still low price ($Cost$ has only increased
from $0.002$ to $0.004$). Qualitatively similar results have been
obtained for other underground systems \cite{lm6}. 
Transportation systems can of course also be analyzed at different 
scales: a similar analysis on a wider transportation system, 
consisting of all the main airplane and highway connections
throughout the world, shows a small-world behavior \cite{lm6}.
This can be explained by the fact that in such a system we 
consider almost all the reasonable transportation alternatives 
available at that scale. In this way the system is closed, 
i.e. there are no other reasonable routing alternatives, 
and so fault-tolerance comes back, after the cost, as a 
leading construction principle.

\section{CONCLUSIONS}
\label{conclusions}

The small-world concept has shown to have lot of appeal both in
sociology (where it comes from), and in science (after the seminal
paper \cite{watts}, a lot of attention has been devoted to this
subject). On the other hand, some aspects of the small worlds were
still not well understood. What is the significance of the
variables involved? Are they ad-hoc parameters, with their somehow
intuitive meaning, or there is a deeper plot? And more: is the
small-world just an abstract concept, applicable in social
sciences or in toy topological models, or does in fact have some
solid grounding in real networks, and can be used in practice to help
us to better understand how real networks work? In this paper we
have tried to cast some light on the above points. We have shown
that already in the topological abstraction, the WS formulation of
the small-world does not work adequately in all cases: because of the
excessive constraints imposed by the formulation, and because of
plain failure to appropriately capture the behaviour of some
networks. 

Therefore it arises the need for a reformulation of the small-world 
concept, which is able to overcome the limitation of the original WS
formulation in the topological abstraction, and also to deal with
the more complex cases of weighted networks. 
The key realization that small-world networks of interest 
represent parallel system, and not just sequential ones, 
brings then to the introduction of
efficiency as the generalizing notion, able to capture the
essential characteristics of the small-world. Efficiency can be
seen as the leading trail that is present both at local and global
level, and allows a smooth extension of the small-world from the
abstractions of the topological world, to the real world of
weighted networks. Together with efficiency, the need for a new
variable also arise by the observation that in real networks, the
target principles of construction (efficiency) also have to take
into account the fact that resources are not unlimited (like in
model 2), and therefore in reality networks have to somehow be a
compromise between the search for performance, and the need for
economy. This new parameter (the cost of a network) nicely couples
with efficiency to provide a meaningful description of the "good"
behaviour of a network, what is called in the paper an {\em
economic small-world\/}. We have shown how local efficiency,
global efficiency and cost can exhibit somehow complex
interactions in dynamically evolving networks, so showing that
economic small-worlds in nature are not trivial to construct and
analyze, but are in fact the product of careful balancing among
these three components. Moreover, the use of these three parameters
also allows a precise quantitative analysis of a network, giving
precise measurements as far as the information flow, and use of
resources, are concerned. So, they give a general measure that can
be used to help us understand not only whether a network is an
economic small world or not, but also to quantitatively capture
with finer degree how these three aspects contribute in the
overall architecture. Finally, we have applied the measures to a
variety of networks, ranging from neural networks, to social
networks, to communication networks, to transportation systems. 
In all these cases, but one, we have seen the appeareance of
the economic small-world behaviour, and even more, we have been
able to push the analysis further, showing in a sense how the
construction principles have played their subtle game of
interaction. 
Moreover, we have shown that the only case of failure of
the economic small-world behaviour (the MBTA), is in a sense just
apparent, and can be explained as the lack of an important, but
often forgotten, underlying feature: the closure of the system.

Summing up, the presented theory seems to substantiate the idea
that efficiency and economy (i.e., economic small-worlds) are the
leading construction principles of real networks. And, the ways
these principles interact can be quantitatively analyzed, in order
to provide us with better intuition on how things work, and how
particular networks better adapt to their specific needs.
\bigskip

{\bf Acknowledgments} We thank A.-L. Barab\'asi, M. Baranger, T.
Berners-Lee, M.E.J. Newman, G. Politi, A. Rapisarda, C. Tsallis
for their useful comments. The Horvitz
Lab at the MIT Department of Biology for fruitful information on
{\em C.~elegans\/} and Brett Tjaden for providing us with the
Internet Movie Database.



%
%


\begin{thebibliography}{99}
\bibitem{algorithm}
Throughout the models and all the real networks presented in this
paper the matrix $\{d_{ij}\}_{i,j\in G}$ 
has been computed by using two different methods:
the Floyd-Warshall ($O(N^3)$) \cite{pallottino}
and the Dijkstra algorithm ($O(N^2~log N)$)
\protect\cite{newman3}.
\bibitem{algorithmnostro}
The two algorithms of \protect\cite{algorithm} can be
easily extended to a weighted network. In this case
the matrix $\{d_{ij}\}$ is calculated using both the information
in $\{a_{ij}\}$ and in $\{\ell_{ij}\}$.
\end{thebibliography}
\end{document}